\documentclass{article}

 \usepackage[preprint]{neurips_2026}

\usepackage[utf8]{inputenc} 
\usepackage[T1]{fontenc}    
\usepackage{hyperref}       
\usepackage{url}            
\usepackage{booktabs}       
\usepackage{amsfonts}       
\usepackage{nicefrac}       
\usepackage{microtype}      
\usepackage{xcolor}         
\usepackage{graphicx}
\usepackage{subcaption}
\usepackage{multirow}
\usepackage{caption}
\title{ CHORDONOMICON: A Dataset of 666,000 Songs and their Chord Progressions }

\author{
Spyridon Kantarelis \\
\texttt{spyroskanta@ails.ece.ntua.gr}
\And
Ioannis Liolitsas \\
\texttt{gliolits@gmail.com}
\And
Konstantinos Thomas \\
\texttt{kthomas@ails.ece.ntua.gr}
\And
Vassilis Lyberatos \\
\texttt{vaslyb@ails.ece.ntua.gr}
\And
Edmund Dervakos \\
\texttt{eddiedervakos@islab.ntua.gr}
\And
Giorgos Stamou \\
\texttt{gstam@cs.ntua.gr}
\AND
Artificial Intelligence and Learning Systems Laboratory \\
National Technical University of Athens, Athens, Greece
}
\begin{document}

\maketitle

\begin{abstract}
Chord progressions encapsulate important information about music, pertaining to its structure and conveyed emotions. They serve as the backbone of musical composition, and in many cases, they are the sole information required for a musician to play along and follow the music. Despite their importance, chord progressions as a data domain remain underexplored — existing datasets lack the scale, structural annotation, and metadata diversity required for rigorous evaluation of music understanding models. In this work, we present \textsl{Chordonomicon}, the largest dataset of its kind, containing over 666,000 song-level symbolic chord progressions, annotated with structural parts (verse, chorus, bridge, etc.), genre, and release date, created by scraping various sources of user-generated progressions and associated metadata, showing strong similarity to well-established prior datasets. Beyond the dataset itself, we propose a reproducible benchmark suite for next chord prediction, evaluating three sequence modeling architectures (RNN, GRU, LSTM) across multiple context window sizes and data scales under strict exact-match evaluation. Our experiments reveal that structural part annotations consistently improve prediction performance. \textsl{Chordonomicon} is released as an open benchmark, providing split methodology, baselines, and evaluation protocols to enable fair and reproducible comparison for future work on chord prediction, classification, generation, and beyond.
\end{abstract}

\begin{figure}[!h]
    \centering
    \includegraphics[scale=0.19]{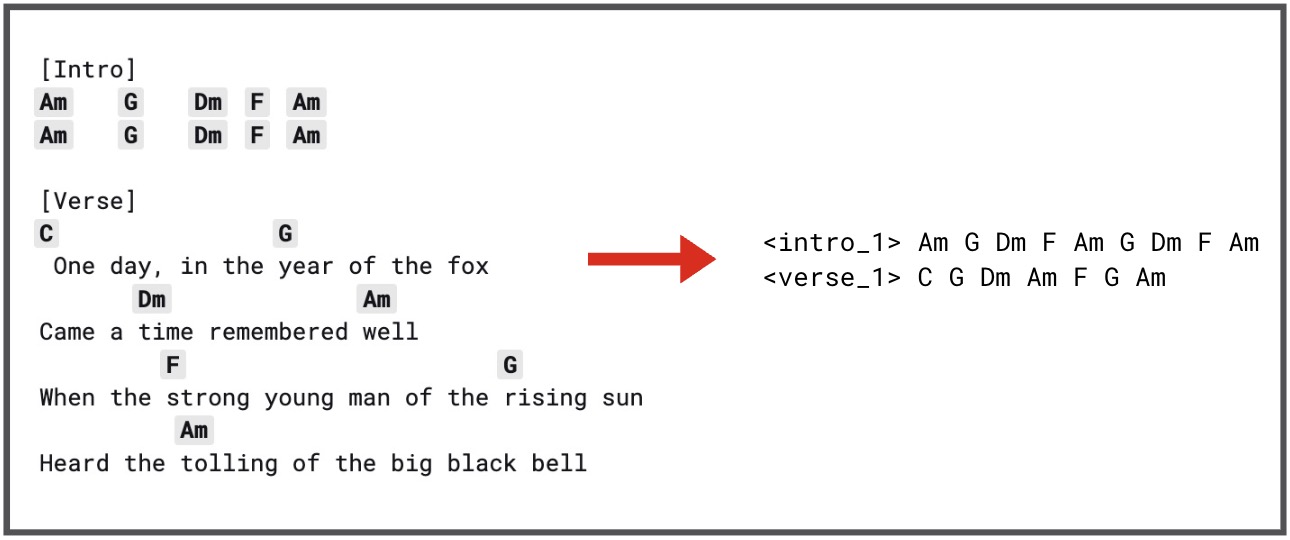}
    \caption{Chords to chord progressions.}
    \label{fig:transition}
\end{figure}

\section{Introduction}\label{sec:introduction}

Large, publicly available, well structured datasets have been crucial for technological progress in recent decades. They allow for wide-spread experimentation and serve as a common basis for comparison, reproducibility, and evaluation. In the domain of music, the development of such datasets has been extremely slow compared to other domains, hindered by issues such as copyrights, unavailability of audio,  lack of symbolic representations, and inadequate metadata. In the era of large language models and text-driven technologies, chord progression datasets have become especially valuable, as they provide structured, text-based representations of music that are easy to process and integrate into modern AI workflows. Among various symbolic formats, chord progressions best capture the harmonic structure of a song—the foundation upon which melodies are constructed and a central element in musical composition. Datasets with significant influence on Music Information Retrieval (MIR) studies, featuring contemporary music data (from the 20th and 21st century) instead of public domain content, have addressed copyright concerns by offering signal-processing attributes rather than copyrighted audio~\cite{bertin} or opting for royalty-free music suitable for commercial purposes~\cite{bogdanov2019mtg}. 

We address these limitations with \textsl{Chordonomicon}~\footnote{\url{https://huggingface.co/datasets/ailsntua/Chordonomicon}}, a novel large-scale dataset of over 666,000 song-level curated non-copyrightable~\cite{booth2016backing} user-generated chord progressions, more than twenty times the amount of chord progression than the largest dataset to date~\cite{de2023choco}, detailed structural annotations, genre labels, release dates, and Spotify IDs. Unlike most previous resources, \textsl{Chordonomicon} provides comprehensive, consistently formatted data suitable for a wide range of machine learning and musicological studies. 

Music datasets often face challenges due to the ambiguity and diverse representations of musical elements~\cite{9222034,article1}, such as chords~\cite{hentschel2021towards}, necessitating curation for semantic integration, as seen in the ChoCo dataset~\cite{de2023choco}, which harmonizes data from 18 sources using heterogeneous formats. To address this, various ontologies have been developed, including the Music Ontology, which provides a framework for music-related information on the Semantic Web~\cite{rashid2018music}, and specialized ontologies like the Music Theory Ontology (MTO)~\cite{rashid2018music}, Functional Harmony Ontology (FHO)~\cite{kantarelis2023functional}, Chord Ontology~\cite{suttonchord}, and Music Part Ontology (MPO)~\cite{kantarelis2023music}, which offer vocabularies and axioms for describing chords, music theory, harmony, and structural analysis. Our methodology aligns with these efforts by representing chords in a way that ensures interlinkability with the Chord Ontology, FHO, and MPO, leveraging their semantic frameworks to enhance dataset integration and analysis, facilitating hybrid AI research that combines symbolic reasoning with data-driven approaches. Chord representations in our dataset follow also the standardized Harte syntax~\cite{harte2005symbolic}, further ensuring consistency and interoperability.


Numerous music datasets have been introduced, incorporating a mix of various formats and metadata that furnishes additional details about the musical pieces within a dataset, encompassing aspects such as music annotations~\cite{speck2011comparative}, genre~\cite{bogdanov2019acousticbrainz}, mood~\cite{bogdanov2019mtg}, artists~\cite{oramas2015semantic}, emotions~\cite{lyberatos2023employing}, and other musical attributes. Such datasets are utilized to address diverse tasks, including but not limited to genre classification~\cite{dervakos2021genre}, music generation~\cite{wang2020pop909} and analysis~\cite{de2011harmtrace}, chord recognition~\cite{nadar2019towards}, and more\footnote{\url{https://www.music-ir.org/mirex/wiki/MIREX\_HOME}}. Chordonomicon is particularly suited for next chord prediction~\cite{lahnala2021chord, korzeniowski2018large} — a symbolic, audio-free task at the intersection of sequence modeling and music understanding. We propose a reproducible benchmark for this task, evaluating multiple architectures and context window sizes, and demonstrate how structural part annotations significantly impact prediction performance. More specifically, in relation to Music Structural Analysis (MSA), there is a scarcity of datasets containing semantic labels~\cite{9747252} necessary for accurately defining distinct music components such as \textit{Verse} and \textit{Chorus}. It is notable that efforts in this field~\cite{paulus2009labelling,smith2011design,nieto2019harmonix} primarily emerged more than a decade ago, and the quantity of semantically labeled music tracks remains notably limited (around 900-1500). In contrast, \textsl{Chordonomicon} offers 2,670,457 structural labels for 397,580 music tracks, adhering to the terminology proposed in these earlier studies.

In this work, we introduce \textsl{Chordonomicon}, outline its data collection and curation pipeline, and discuss its representational versatility. We propose a reproducible benchmark suite for next chord prediction, establishing baseline performance across four architectures, multiple context window sizes, and data scales under strict exact-match evaluation. Our experiments yield concrete findings: structural part annotations provide significant performance gains, the optimal harmonic context window is approximately 10--20 chords, and LSTM outperforms more recent architectures — collectively suggesting that next chord prediction is governed by local harmonic dependencies rather than long-range sequential structure. We release all splits, baselines, and evaluation protocols openly, with the goal of establishing \textsl{Chordonomicon} as a standard benchmark for reproducible and comparable research in computational harmony. While the proposed benchmark is intentionally simple, we view it as a starting point for more comprehensive evaluations on symbolic harmony modeling.

\section{Dataset}

Chordonomicon is a symbolic dataset containing exclusively chord progressions and associated metadata — no audio, lyrics, or song titles are included, as these constitute copyrightable material unsuitable for open research datasets. This section delves into insights related to data collection and processing. Specifically, we provide detailed descriptions of the methodology used for data collection and curation, while at the same time explaining the process of data selection and exclusion. Table \ref{tab:statistics} summarizes the dataset's key statistics. On average, each progression consists of 76.48 chords, with a median of 71 chords and a standard deviation of 54.59.

\subsection{Data Gathering and Curation}

The \textsl{Chordonomicon} dataset was created by carefully collecting, processing, and standardizing data to ensure high-quality, uniform song-level chord progressions and metadata suitable for machine learning (see \ref{fig:transition}). We leveraged web scraping techniques to extract song chord data from the Ultimate Guitar (UG) platform. By adhering to UG's robots.txt, Terms of Service, and established web accessibility protocols (see \ref{sec:ethical}), we obtained the site's XML sitemap and filtered the URLs containing the ``\emph{-chords-}'' substring, which denotes song chord pages. This yielded a dataset of 1.4 million song chord pages, encompassing multiple interpretations (versions) per song. We then collected the raw content, user ratings, vote counts, artist IDs, song IDs, artist names, and song titles from each of these URLs. Utilizing the song and artist IDs, along with fuzzy string matching on the song and artist names, we consolidated this data into 793,362 unique songs from 93,872 unique artists. Duplicate songs were consolidated using Levenshtein distance-based fuzzy matching with a similarity threshold of 0.85, handling common user-generated inconsistencies such as missing articles, version annotations, and spelling errors (e.g., "Hotel California" vs. "The Hotel California"). For songs with multiple chord versions, we retained the best-rated version, determined by a weighted average of user rating and vote count percentage. The raw HTML files were parsed to extract full chord progressions for each song. To ensure consistency, consecutive identical chords were collapsed into a single token (e.g., 'A G G A' → 'A G A'), as repetition carries no additional harmonic information. Songs with fewer than four chords were excluded as they lack sufficient harmonic content for meaningful analysis. All files were standardized to UTF-8 encoding, with any songs containing non-standard or non-Latin chord encodings (e.g., Cyrillic characters) removed to ensure uniform representation. All associated metadata were transformed into a unified format to eliminate cross-source ambiguities.

The remaining dataset underwent expert curation over a period of approximately three months by two specialists holding advanced degrees in Western tonal harmony. Western tonal music provides a well-established, theoretical framework defining all chords derivable from major, harmonic minor, and melodic minor scales, offering an unambiguous criterion for harmonic validity. For each chord, experts verified: (i) the presence of an identifiable tonic note (A through G), (ii) the correctness of any alterations (sharps and flats), and (iii) that the chord quality is consistent with Western tonal structures. Progressions containing notationally invalid symbols — such as entries lacking a tonic note entirely (e.g., "add\#9"), numerals misused as chords (e.g., "12"), or locale-specific notations incompatible with standard Western notation (e.g., "fis") — were removed entirely. This process ensured that every chord in Chordonomicon is unambiguous, theoretically grounded, and convertible to standardized machine-readable formats. This rigorous curation process resulted in the removal of over 100,000 songs — approximately 13\% of the consolidated dataset — yielding a final collection of 679,807 unique, fully curated song-level chord progressions, harmonically accurate, semantically rich, and convertible to standardized machine-readable formats.

All chord progressions are provided in the FHO syntax, with conversion scripts provided for the Harte chord syntax — the de facto standard for chord representation in MIR, widely adopted in prominent datasets such as the McGill Billboard corpus~\cite{burgoyne2011expert}. Crucially, the successful conversion of every chord in Chordonomicon to Harte syntax serves as an additional validation of dataset quality — every entry, including locale-specific and non-standard notations encountered during curation, can be unambiguously mapped to a standard Harte representation (e.g., "SOL5" → "G:(3)", "hmin" → "B:min"), formally confirming that the dataset contains exclusively well-formed, theoretically grounded harmonic content.

Structural part annotations are a defining feature of Chordonomicon and, as our experiments demonstrate, a critical factor in chord prediction performance. We employed eight distinct part categories: \textit{Intro, Verse, Chorus, Bridge, Interlude, Solo, Instrumental}, and \textit{Outro}, resolving approximately 3,300 unique user-generated part name variants to their canonical category (e.g., "Refrain" → "Chorus", "Verse 1" → "Verse"). Songs containing any unclassifiable part label were excluded from structural annotation, retaining only their chord progression — a conservative approach that ensures reliability: 397,580 songs (58.5\% of the dataset) carry full structural annotations. Release dates and genre metadata were collected via the Spotify Web API, verified by matching track and artist names across both sources, with Spotify IDs provided where available.


\subsection{Genres}

\begin{figure}[htbp]
    \centering
    \includegraphics[scale=0.12]{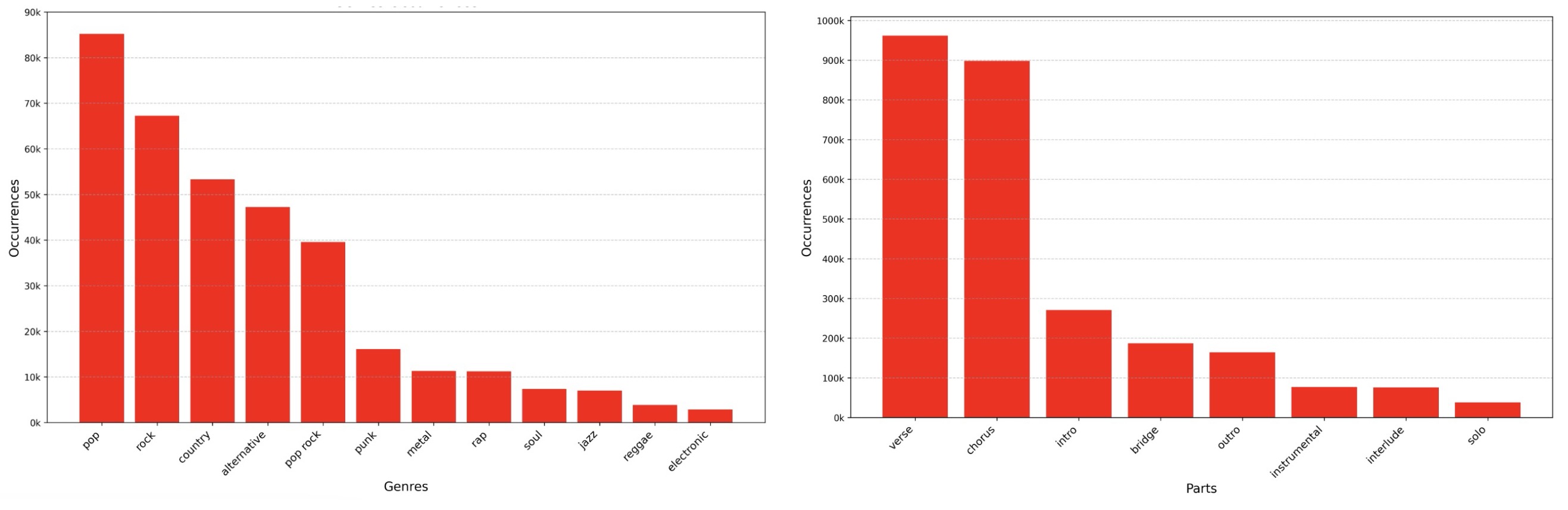}
    \caption{Dataset genres' and parts' distribution.}
    \label{fig:genres_parts}
\end{figure}

Spotify encompasses genres that can be quite specific and may pose challenges for users to easily comprehend (e.g., \textit{lilith}, \textit{otacore}). To address this complexity, we categorized the genres into the 12 most prevalent ones (see Figure~\ref{fig:genres_parts}) following an analysis of our data and referring to the genre genealogy provided by Musicmap\footnote{\url{https://musicmap.info}}. Importantly, while Chordonomicon employs Western tonal notation as its representational framework, this does not limit its cultural scope. Through Spotify's genre metadata, the dataset encompasses music from a wide range of geographic and cultural origins, including Latin American music, Japanese popular music (\textit{otacore}), South Asian traditions (\textit{rabindra sangeet}), African genres, Caribbean music (\textit{reggae}), and many others — all of which operate within or alongside Western harmonic frameworks. We encoded release dates by decade, resulting in 15 classes spanning from the 1890s to the 2020s, with the majority of tracks from the 2010s onward, reflecting the contemporary focus of the Ultimate Guitar platform. This makes Chordonomicon a resource for studying not only Western popular music but also the global spread and adaptation of Western tonal harmony across cultures and decades.

\begin{figure}[ht]
\centering

\begin{minipage}{0.48\textwidth}
\centering
\refstepcounter{table}
\tablename~\thetable: Dataset Key Statistics.
\label{tab:statistics}

\vspace{0.5em}

\renewcommand{\arraystretch}{1.1}
\begin{tabular}{|l|r|}
\hline
\textbf{Data Property} & \textbf{Occurrences}\\
\hline\hline
No. of Tracks & 679,807\\
\hline
No. of Chords & 51,994,634\\
\hline
No. of Unique Chords & 749\\
\hline
No. of Unique Inverted Chords & 3,577\\
\hline
No. of Tracks with Parts & 397,580\\
\hline
No. of Parts & 2,670,457\\
\hline
No. of Tracks with Release Date & 422,181\\
\hline
No. of Tracks with Spotify Genres & 429,753\\
\hline
No. of Tracks with Main Genre & 352,111\\
\hline
No. of Tracks with Rock Sub-Genre & 145,218\\
\hline
No. of Tracks with Spotify Track ID & 440,284\\
\hline
No. of Tracks with Spotify Artist ID & 510,986\\
\hline
\end{tabular}

\label{tab:statistics}
\end{minipage}
\hfill
\begin{minipage}{0.33\textwidth}
\centering
\includegraphics[width=\linewidth]{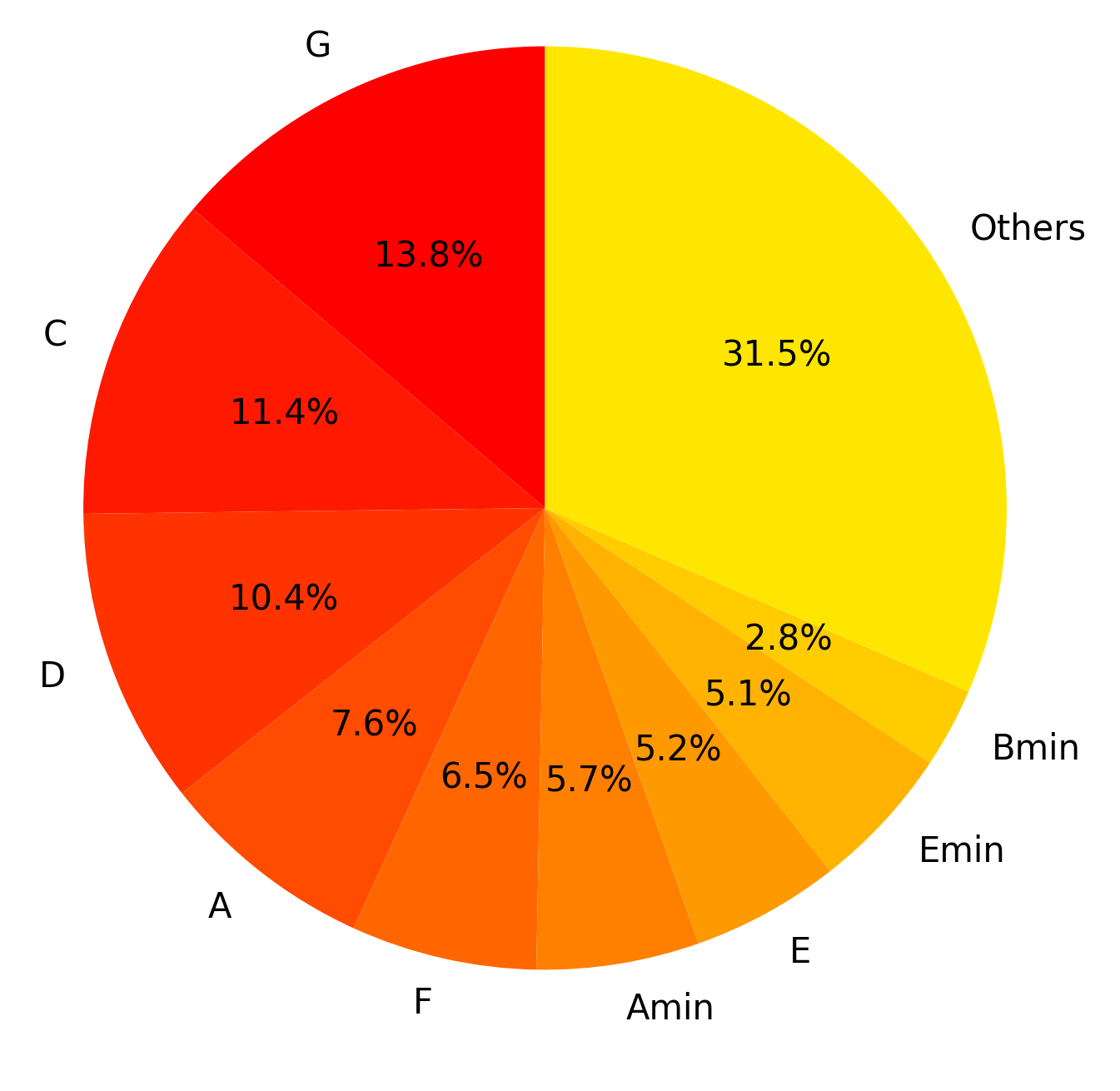}
\caption{Dataset chord distribution.}
\label{fig:chords_distribution}
\end{minipage}

\end{figure}

\subsection{Chords}

A music chord is a group of notes played together. In Western music, chords are typically built on intervals of thirds (tertian harmony), with triads (three notes) forming the foundation and extended chords comprising four to seven notes. An exception is the power chord (two notes a fifth apart) commonly employed in rock and metal music. When the notes of a chord are reordered, placing a non-root note in the bass, the result is an inverted chord. Our dataset comprises 51,994,634 chord tokens, with 749 unique chord types and 1,464,392 inverted chord instances. 

The chord distribution, illustrated in Figure~\ref{fig:chords_distribution} and closely mirroring the ground-truth reference distributions of established chord recognition datasets~\cite{burgoyne2011expert}, reveals a clear dominance of natural triads, particularly chords in the C and G major families. This is not a sampling artifact but a well-understood phenomenon musicologically. First, G, C, D, and A are open chord shapes on guitar, requiring minimal finger placement and making them the default vocabulary of beginner and intermediate players. Second, the widespread use of capos systematically shifts transcriptions toward guitar-friendly positions regardless of the song's original tonality. Since Chordonomicon is sourced from a guitar-oriented platform, this distribution authentically reflects how popular music is actually performed and transcribed by guitarists worldwide, rather than how it is theoretically composed. Additionally, non-professional transcribers often simplify complex harmonies to their triad equivalents, contributing to the dominance of simple chord forms. 

\begin{figure}[]
    \centering
    \includegraphics[scale=0.45]{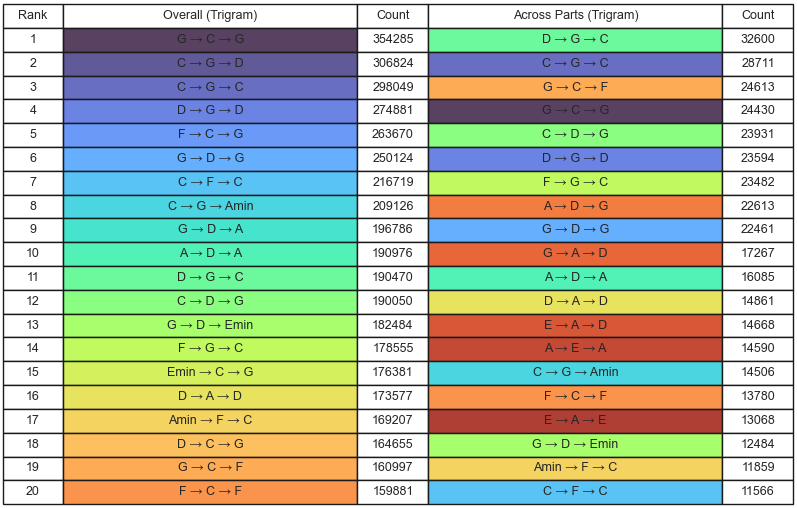}
    \caption{Top 20 Chord Trigrams Overall vs Across Parts.}
    \label{fig:trigrams}
\end{figure}

Figure~\ref{fig:trigrams} presents the top 20 chord trigrams overall and across part boundaries, where the final chord of the trigram belongs to a different structural part than the preceding two. The two distributions differ, revealing that chord transition probabilities shift depending on structural context, suggesting that part annotations carry meaningful predictive information for next chord prediction. In particular, the first place ($G$-$C$-$G$) is taken over by the 11th ($D$-$G$-$C$) one when switching from overall to across parts, with the previous last place becoming 4th, while a good amount of the top 20 trigrams overall do not make the top 20 between part boundaries.

\section{Next Chord Prediction Benchmark}

Next chord prediction is a task where given an incomplete chord progression, we have to predict the next chord in the sequence. Formally, given a sequence of chords $(y_1,y_2,\dots,y_k)$ we predict $P(y_k|(y_1,\dots,y_{k-1})$, which means that this task is essentially language modeling. We focus on next chord prediction as a canonical and well-defined task that enables controlled evaluation across models and data scales. While the dataset supports a broader range of tasks (e.g., generation, classification, and structure prediction), we leave their systematic benchmarking to future work. The experimental framework was designed to assess how training set size, sequence length, chord representation, and structural information influence model performances. The code for replicating all benchmark experiments is publicly available at \url{https://github.com/spyroskantarelis/chordonomicon}.

\subsection{Experimental Framework}

For the chord representation, we used two approaches: a \textbf{single-token} representation, where each unique chord symbol (e.g., Cmaj7, Am, G/B) was treated as an atomic token, and a \textbf{three-tokens} representation, where each chord was decomposed into three components. More specifically the Root: the pitch class of the chord (e.g., C, Eb); Quality + Extensions: the chord type and added tones (e.g., maj7, min, sus4, add13); and Bass: the bass note when specified (e.g., the E in C/E), or N for none. For the structural information, we created three large subsets of our dataset: \textbf{Filtered dataset:} Contains all songs with at least one metadata field (genre or decade), with structural labels removed; suitable for evaluation on entire songs (472483 rows). \textbf{Labeled-song dataset:} Subset of the filtered dataset containing only songs with structural labels (296179 rows), later kept without labels for evaluation. \textbf{Segmented dataset:} Each entry corresponds to a single labeled section (e.g., a single verse or chorus - 2009699 rows). Labels and associated chord sequences are preserved for section-level modeling and evaluation.

For the tasks three architectures were considered, the vanilla recurrent neural networks (RNN), the gated recurrent units (GRU), the long short-term memory networks (LSTM). Transformer-based architectures are left for future work, as their effective application to chord sequences would require investigation of alternative tokenization strategies and data representations suited to the limited harmonic vocabulary. The models receive sequences of either encoded chords (1-token) or encoded chord components (3-tokens) and uses an embeddings layer to create meaningful embeddings to be processed by the main module. The module's output is then collected by fully connected layers, with either one head per chord or 3 heads, one per chord component. For the 3-tokens representation the heads of the components are then combined to determine which chord from the known vocabulary fits best. The multi-head approach allows the model to leverage both component-level and full-chord supervision simultaneously.  

Models were trained using cross-entropy loss, with the total loss of the 3-tokens approach being the summation of the component and full chord losses. Hyperparameters were as follows: hidden dimension 256, one main module layer, batch size of 4096, learning rate 0.005, a maximum of 200 epochs and embedding dimension of 128 for the chords and 16 for each component, resulting to approximately 449k--1.3m parameters depending on model and tokenization. Training was performed on the segmented dataset, while evaluation was performed on the filtered dataset, the labeled-song dataset, and the segmented dataset after filtering songs that had at least one part belong to the training or validation dataset. To improve performance and generalization, added mechanisms were employed. Early stopping in combination with learning rate scheduling was used to avoid overfitting, while a stratified splitting ensured training validation and test sets had no overlapping songs and maintained label distributions. Also, during all phases, to increase robustness to variable-length sequences, sequences were cut randomly when sampling while ensuring the target chord followed at least 4 chords. All experiments were conducted locally on a consumer-grade laptop equipped with an Intel i7-9750H CPU and an NVIDIA GTX 1650 GPU, with each training run lasting between 10 minutes and 2 hours. Across approximately 270 training runs, result variance remained consistently below 1\%, confirming statistical stability. We estimate total energy consumption at approximately 9--18 kWh, corresponding to a carbon footprint of roughly 3.5--9 kg CO$_2$ based on an average grid carbon intensity of 300--500g CO$_2$/kWh.

\subsection{Results}

\begin{figure}[htbp]
    \centering
    
    \begin{subfigure}[t]{0.44\textwidth}
        \vspace{0pt}
        \centering
        \includegraphics[width=\linewidth]{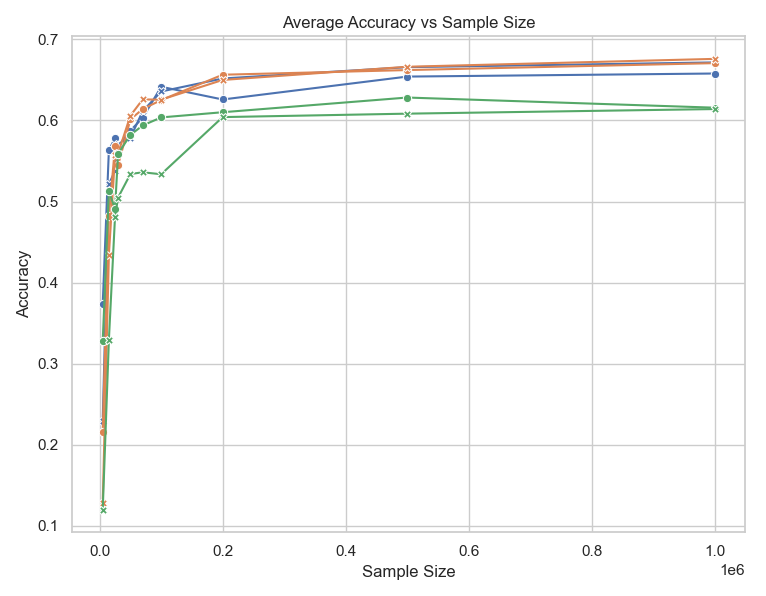}
        \caption{Average Cumulative Accuracy}
    \end{subfigure}
    \hfill
    \begin{subfigure}[t]{0.44\textwidth}
        \vspace{0pt}
        \centering
        \includegraphics[width=\linewidth]{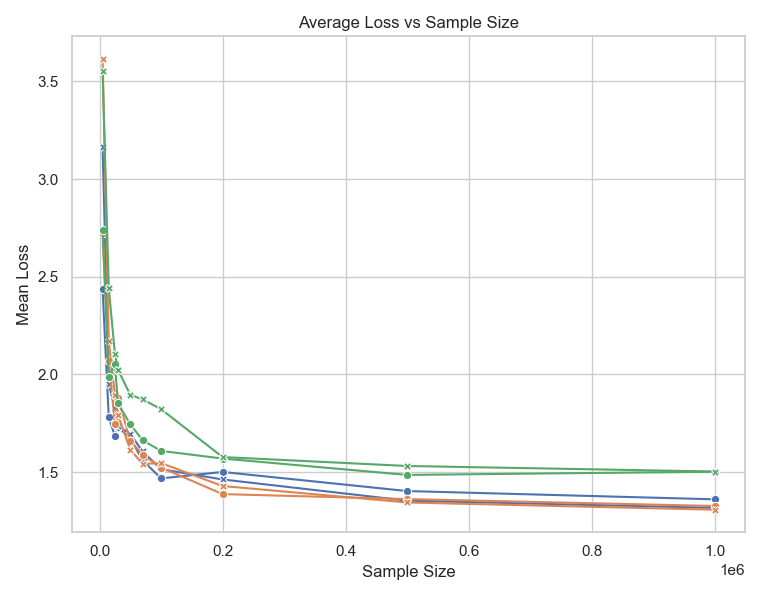}
        \caption{Average Cumulative Loss}
    \end{subfigure}
    \hfill
    \begin{subfigure}[t]{0.1\textwidth}
        \vspace{7.6pt}
        \centering
        \includegraphics[width=\linewidth]{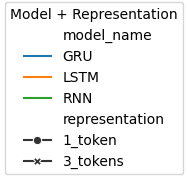}
    \end{subfigure}
    
    \caption{Average Cumulative Accuracy and Loss for the Segmented subset vs Sample Size.}
    \label{fig:sample_size}
\end{figure}

The chord prediction experiments consist of two main analyses. First, we investigated how the size of the training dataset influences model performance, as shown in Figure \ref{fig:sample_size}. Performance improves dramatically as the sample size grown from a few thousand entries (specifically 5,000) to 200,000, with only minor gains beyond that up to 1,000,000 samples. RNN consistently performs worse than the other models. Table \ref{table:rnn_summary} summarizes the average results over multiple runs using the best performing sample size (1,000,000 samples). When tested with whole song sequences that have part labels, accuracy decreases and loss increases. Performance drops further when tested on datasets missing structural labels. The 3-tokens representation generally outperforms the 1-token representation, except for the vanilla RNN. In terms of accuracy, the LSTM achieves the best results, while the GRU shows slightly better performance in loss despite a minor gap in accuracy. This suggests that next chord prediction in this setting is dominated by strong local dependencies and relatively low-entropy transitions, where the inductive biases of recurrent models remain competitive or advantageous.

\begin{figure}[htbp]
    \centering
    
    \begin{subfigure}[t]{0.28\textwidth}
        \vspace{0pt}
        \centering
        \includegraphics[width=\linewidth]{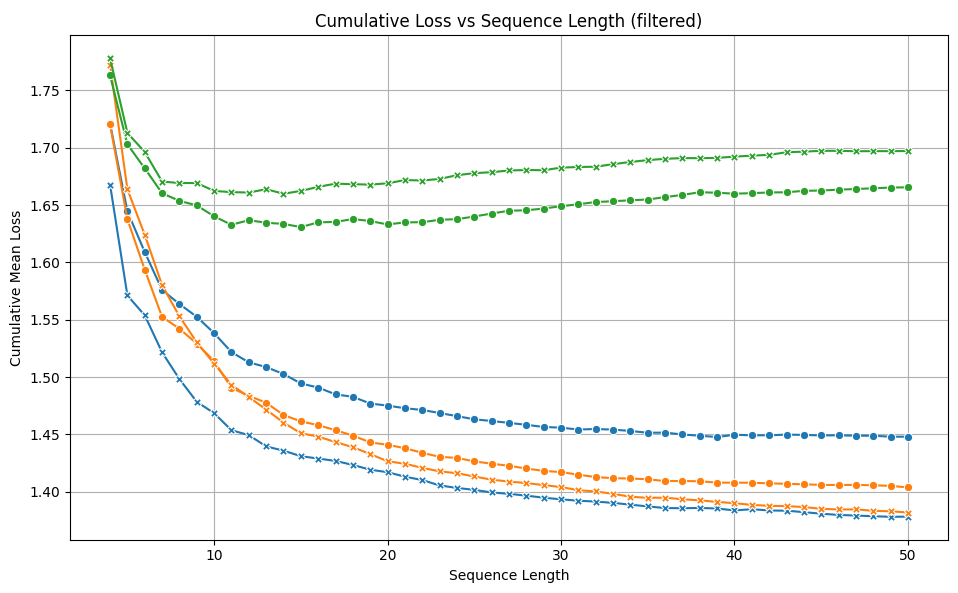}
        \caption{Filtered subset}
    \end{subfigure}
    \hfill
    \begin{subfigure}[t]{0.28\textwidth}
        \vspace{0pt}
        \centering
        \includegraphics[width=\linewidth]{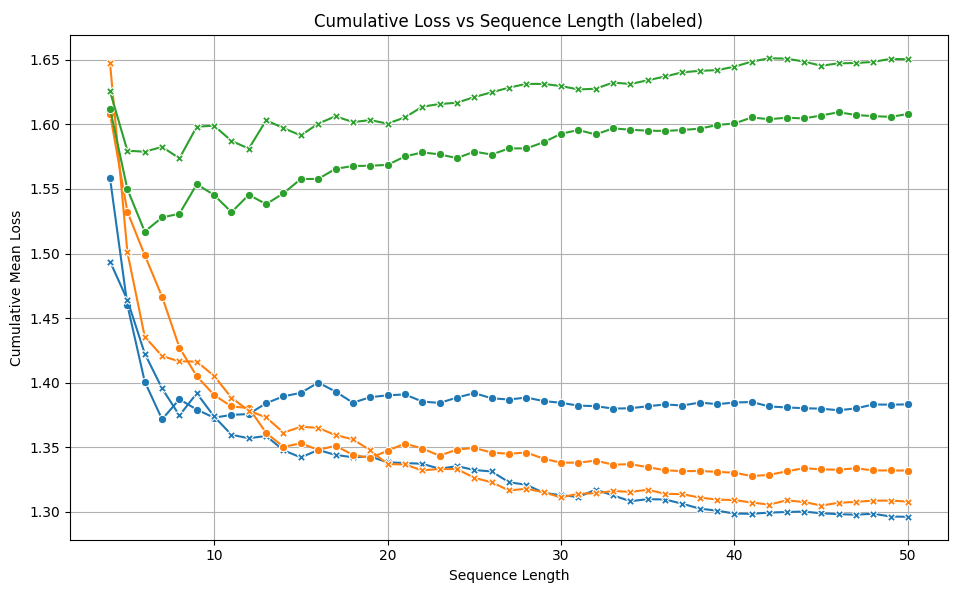}
        \caption{Labeled subset}
    \end{subfigure}
    \hfill
    \begin{subfigure}[t]{0.28\textwidth}
        \vspace{0pt}
        \centering
        \includegraphics[width=\linewidth]{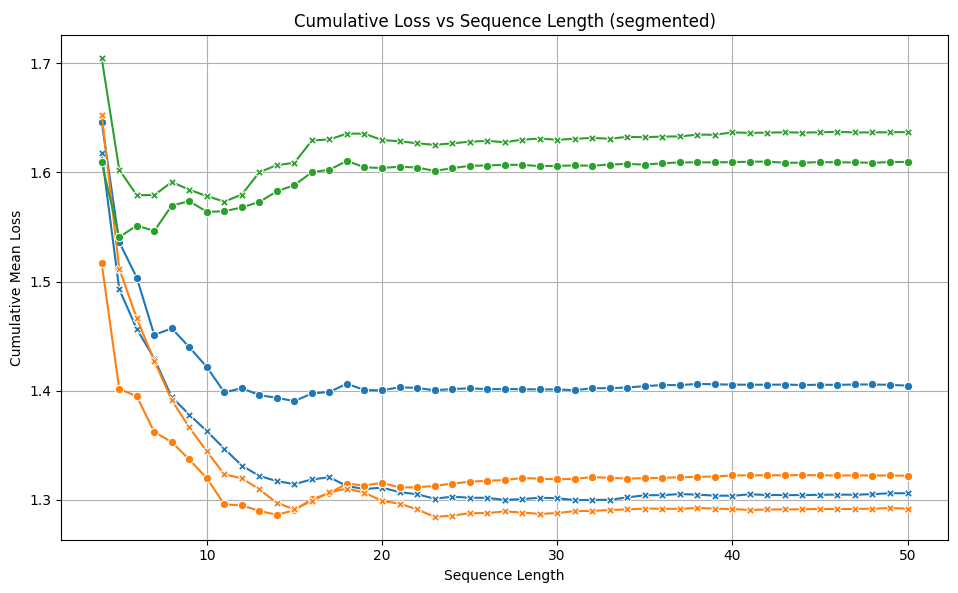}
        \caption{Segmented subset}
    \end{subfigure}
    \hfill
    \begin{subfigure}[t]{0.1\textwidth}
        \vspace{3.6pt}
        \centering
        \includegraphics[width=\linewidth]{figs/captions.png}
    \end{subfigure}
    
    \caption{Cumulative Loss for all subsets vs Sequence Length.}
    \label{fig:sequence_length}
\end{figure}

Secondly, we investigated the impact of structural information and sequence length on next chord prediction by evaluating all architectures across the three dataset subsets. Figure~\ref{fig:sequence_length} presents the cumulative mean loss versus sequence length across all conditions. LSTM consistently achieves the lowest loss while RNN clearly underperforms, with GRU sitting between the two — particularly benefiting from the 3-token representation over the 1-token approach. Performance stabilizes at approximately 10--20 chords of context, suggesting this as the effective context window for next chord prediction. Most notably, loss decreases progressively from the filtered subset to the labeled subset to the segmented subset, demonstrating that structural information consistently reduces prediction uncertainty across all architectures and representation choices. Our findings agree with those in~\cite{korzeniowski2018large}, confirming both the relative ordering of model performances and the effective context window of approximately 10--20 chords.

\subsection{Discussion}

Across all architectures, performance improves steeply with training data size until approximately 200,000 samples, after which gains plateau, indicating that the dominant harmonic patterns in the corpus are captured well before exhausting the full dataset, with additional samples providing diminishing returns. A similar saturation effect is observed with sequence length, where performance stabilizes at approximately 10--20 chords, confirming that harmonic predictability is concentrated locally. The segmented subset directly exploits this locality; by treating each structural part as an independent sequence, models operate within a harmonically coherent context window, which explains its consistent performance advantage over the labeled and filtered subsets. The 3-token representation consistently outperforms the 1-token approach across all conditions, as decomposing each chord into its root, quality, and inversion allows models to internalize relationships between chord components.

When models are evaluated on entire songs without part segmentation, overall performance degrades. 
Interestingly, although accuracy declines, loss continues to decrease, suggesting that improvements in the objective may reflect increased confidence or calibration effects rather than gains in classification accuracy.
This occurs because crossing a part boundary effectively resets the harmonic context: the tonal patterns governing one section do not carry over to the next, and sequences spanning multiple parts blend incompatible harmonic contexts. The gradual confidence gain reflects the model adapting to the new local context after the boundary, rather than genuine predictive improvement. This reinforces the importance of part-level structure in symbolic chord modeling.

\begin{table}
\centering
\caption{Average Accuracy and Loss per Model and Representation across Datasets. Best results are in \textbf{bold}, second best are marked with $^{*}$.}
\makebox[\textwidth]{%
\begin{tabular}{ll|cc|cc|cc}
\hline
 & & \multicolumn{2}{c|}{\textbf{Segmented}} & \multicolumn{2}{c|}{\textbf{Labeled}} & \multicolumn{2}{c}{\textbf{Filtered}}\\ 
\cline{3-8}
\textbf{Model} & \textbf{Representation} & \textbf{Avg Acc} & \textbf{Avg Loss} & \textbf{Avg Acc} & \textbf{Avg Loss} & \textbf{Avg Acc} & \textbf{Avg Loss} \\ 
\hline

\multirow{2}{*}{\textbf{RNN}} 
 & \textbf{1-token} & 0.5859 & 1.6369 & 0.5632 & 1.6081 & 0.5448 & 1.6655 \\
 & \textbf{3-tokens} & 0.5715 & 1.6370 & 0.5481 & 1.6504 & 0.5281 & 1.6972 \\
\hline

\multirow{2}{*}{\textbf{LSTM}} 
 & \textbf{1-token} & 0.6594 & 1.3221 & 0.6244 & 1.3319 & 0.6007 & 1.4037 \\
 & \textbf{3-tokens} & \textbf{0.6748} & \textbf{1.2919} & \textbf{0.6394} & 1.3079$^{*}$ & \textbf{0.6121} & 1.3818$^{*}$ \\
\hline

\multirow{2}{*}{\textbf{GRU}} 
 & \textbf{1-token} & 0.6399 & 1.4046 & 0.6017 & 1.3832 & 0.5791 & 1.4479 \\
 & \textbf{3-tokens} & 0.6618$^{*}$ & 1.3060$^{*}$ & 0.6311$^{*}$ & \textbf{1.2961} & 0.6055$^{*}$ & \textbf{1.3782} \\ 
\hline
\end{tabular}%
}

\label{table:rnn_summary}
\end{table}
\section{Limitations}

Chordonomicon has several limitations inherent to its source. Timing and beat information is absent, as Ultimate Guitar transcriptions do not encode temporal structure. Guitar-idiomatic patterns (simplified voicings and capo usage) are present, reflecting authentic popular music performance practice. As the dataset is derived from user-generated content, it inevitably contains noise and simplified harmonic representations. Rather than attempting to fully eliminate such variability, we view it as reflecting real-world transcription practices. The scale of the dataset helps mitigate individual errors, though we acknowledge that systematic biases (e.g., guitar-centric chord distributions) remain. The equal temperament nature of the guitar means microtonal and non-Western harmonic systems are outside the dataset's scope. Each song's tonal key was excluded as its assignment is a globally interpretive judgment, unlike individual chord validity which is locally verifiable; transposition scripts are provided for key normalization. N.C. (no chord) markers were excluded as they carry no harmonic information.

\section{Conclusion}

We present the \textsl{Chordonomicon}, the largest dataset of chord progressions, comprising over 666,000 meticulously curated user-generated progressions, along with uniformly formatted structural information and additional metadata, addressing critical gaps in data availability and consistency for chord-based MIR tasks. Its standardized formatting and integration potential with music ontologies open new possibilities for hybrid AI systems and advanced harmonic analysis. Additionally, we offer insights into its corpus and demonstrate baseline performance on a proposed evaluation framework for next chord prediction, investigating different model architectures, representations, sample sizes, sequence length, and structural information. Results highlight the positive effect of structural information. 

Finally, \textsl{Chordonomicon} offers a unique blend of harmonic, structural, and metadata-rich representations on a very large scale, making it an invaluable tool for AI researchers across domains. We envision this dataset as a foundation for advancing chord-based generative modeling, classification tasks, and hybrid AI systems. We encourage the research community to build upon our preliminary benchmarks, explore novel architectures, and deepen our understanding of chord progressions in music theory and computational creativity.


\section*{Ethics Statement}\label{sec:ethical}

\paragraph{Data Harvesting Legality and Ethicality}
1. \textit{Web Crawler Permissions}: Ultimate Guitar (hereafter  ``UG") does not flag  "tabs.ultimate-guitar.com/tab/*" as  ``Disallowed'' in their /robots.txt file. This indicates that UG permits automated crawlers to access and index their tablature and chord pages, aligning with established web protocols for content accessibility.
2. \textit{Terms of Service Compliance}: A thorough examination of UG's Terms of Service (TOS) reveals no obvious prohibition against automated content crawling or scraping. This absence of such restrictions further supports the permissibility of data collection for research purposes.
3. \textit{Legal Precedents}: U.S. courts have consistently ruled in favor of web-scraping publicly accessible data, particularly for research purposes. Notably, in Sandvig v. Barr (D.D.C. 2020), the court held that such activities are protected under the First Amendment and do not violate the Computer Fraud and Abuse Act (CFAA). Additionally, the hiQ Labs, Inc. v. LinkedIn Corp. (9th Circuit, 2019) decision further reinforced the legality of scraping publicly available data.
\paragraph{Intellectual Property Considerations}
1. \textit{User-Generated Content Ownership}: UG's TOS explicitly states that  ``Ultimate Guitar does not claim any ownership rights in User Generated Content that you transmit, submit, display or publish on, through or in connection with the Service". And further clarifies that  ``User Generated Content includes, without limitation, tablatures (text or electronic)...". This clause clearly delineates the ownership status of the content we're collecting.
2. \textit{Non-Copyrightable Elements}: Legal precedents consistently support the notion that chord progressions are not copyrightable. Cases such as Granite Music Corp. v. United Artists Corp. (1977) and Swirsky v. Carey (2004) have established that chord progressions are fundamental building blocks of music, akin to "common musical property", since many songs share the same or similar chord progressions. This legal stance corroborates that our dataset solely comprises of non-copyrightable elements.
3. \textit{Data Processing}: To further ensure ethical and legal compliance, our data collection process excludes copyrightable elements such as lyrics, song titles, and artist names. We retain only the chord progressions, the associated song structure information (e.g., intro, verse, chorus), release year, the genres of the relevant artist and Spotify IDs of track and artist (non-copyrightable alphanumeric strings).
\paragraph{Context and Licensing}
1. \textit{Precedent in Existing Datasets}: The Common Crawl dataset, widely used for training most top-tier large language models, consistently crawls and already includes numerous, if not all, chord pages from Ultimate Guitar\footnote{\url{https://www.ultimate-guitar.com/}} in its archives. A relevant search on the dataset\footnote{\url{https://index.commoncrawl.org/}} will show all the harvested UG URLs, and the relevant Common Crawl web archive (.warc) files will show the raw scrapped pages, along with the lyrics, song names, bands, and all information visible on each tab's webpage. This precedent demonstrates the accepted practice of including such data in large-scale research datasets.
2. \textit{Licensing Considerations}: The Common Crawl dataset's license is a limited license that allows users to access and utilize the data while agreeing to respect the copyrights and other applicable rights of third parties in and to the material contained therein. Despite the non-copyrightable nature of our data, we still adopt the same limited licensing approach for peace of mind.

In conclusion, by adhering to web crawling conventions, respecting Terms of Service, focusing on non-copyrightable elements, and following well-established precedents in data collection and licensing, our approach to creating this dataset stands on solid legal and ethical ground.  Beyond legal compliance, we recognize the importance of respecting the creative efforts of the Ultimate Guitar community, whose voluntary transcription work made this dataset possible. While the anonymous nature of user-generated contributions makes individual consent and attribution infeasible, we explicitly acknowledge the UG community's indirect but essential role in enabling this research. Our data collection avoids any personally identifiable information, retaining only non-copyrightable harmonic content, aligning with both legal and ethical norms in data research. The potential scientific and cultural value of this research further justifies its creation and use within the bounds of fair use and academic freedom.

\newpage

\medskip
\bibliographystyle{plain}
\bibliography{neurips}


\end{document}